# Three-phase contact line dynamics on moving fibers measured by X-ray holography

Louisa E. Kraft[1,2], Jens Lucht[3], Fiona Berner[1,2], Hannes P. Hoeppe[3], Tobias Eklund[1,2,4], Yizhi Liu[1], Markus Osterhoff[3], Fabian Westermeier[5], Wojciech Roseker[5], Tim Salditt[3], Hans-Jürgen Butt[1], Katrin Amann-Winkel[1,2,*]

[1]Max Planck Institute for Polymer Research, Ackermannweg 10, 55128 Mainz, Germany

[2]Institute for Physics, Johannes Gutenberg University, Staudingerweg 7, 55128 Mainz, Germany

[3]Institut für Röntgenphysik, Georg-August-Universität Göttingen, Friedrich-Hund-Platz 1, 37077 Göttingen, Germany

[4]European XFEL, Holzkoppel 4, 22869 Schenefeld, Germany

[5]Deutsches Elektronen-Synchrotron, Notkestrasse 85, 22607 Hamburg, Germany

# Abstract

Wetting of solid surfaces by a liquid is important for many natural and industrial processes such as printing, painting and coating. However, a quantitative description of the dynamic receding and advancing contact angle is still debated, in particular for aqueous solutions. One reason for our lack of quantitative understanding is the limited spatial resolution of currently used optical methods. We therefore present a new approach to access the sub-microscopic region. We use X-ray phase contrast imaging to measure the dynamic receding contact angle on a moving glass fiber of 17 μm diameter. The fiber was pulled out of a liquid bath which was filled with a mixture of glycerol and Milli-Q water. The dynamic receding contact angle decreased with increasing contact line velocity for all mixtures. In the holograms we achieved a resolution of 50 nm/pixel with a spatial error of 450 nm. This spatial error is due to an extended surface region of the fiber and the liquid surface in the holograms. Our results demonstrate the feasibility of X-ray holography as a method to investigate dynamic contact angle phenomena and thereby opening pathways to higher spatial and temporal resolution.

# Abstract Graphic

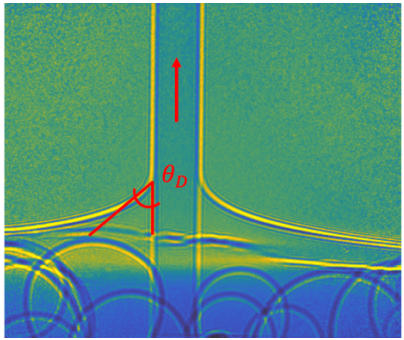

# Introduction

Wetting of solid surfaces is important for many industrial as well as natural processes, especially with respect to dynamical wetting. Examples include coating, painting, inkjet printing, heat exchangers and flotation [1]. When a liquid drop is deposited on a non-wetting or partially wetting surface the contour of the liquid forms a finite contact angle in relation to the surface. A three-phase contact line forms at the interface of the vapor, the liquid, and the solid. Its tangent, the contact angle, is characteristic for the specific combination of vapor, solid and liquid. In equilibrium this contact angle is related to the involved interfacial free energies via Young's equation [2, 3, 4]. Experimentally, however, the equilibrium contact angle as stated in Young's equation does not exist. Further, the solid-vapor and the solid-liquid interfacial energies cannot be measured. Instead, experimentally, a whole range of contact angles is detected for a specific solid-liquid combination even on apparently homogeneous and smooth solids. The actually observed contact angle depends on how the wetting situation was created, e.g. how a drop was placed onto a surface. This range of possible contact angles is bounded by a lower and upper limit. The lower limit is the receding, the upper limit the advancing contact angle [5, 6]. This difference of the advancing and receding contact angle was termed 'contact angle hysteresis' [7]. Factors leading to contact angles hysteresis are roughness [8], heterogeneity [9, 10], adaptation [11], slide electrification [12] and, on very soft surfaces, mechanical deformation [13]. Contact angle hysteresis can therefore nowadays be used to characterize surfaces.

Significant progress has been made in terms of experimental characterization as well as understanding the static and macroscopic properties of liquid drops on solid surfaces like the contact angle [14]. However, certain aspects of the wetting dynamic of solid surfaces are still under debate. For example, the precise dynamic mechanism, with which the contact line advances across a solid can thus far not be described quantitatively. A central observation in dynamic wetting is the velocity dependence of the dynamic advancing and receding contact angles. The dynamic advancing contact angle increases with increasing speed of the contact line, while the dynamic receding contact angle decreases with increasing velocity [15]. Different models have been proposed to describe this velocity dependence. They assume different energy dissipation processes. The hydrodynamic model assumes viscous dissipation as the major source of energy dissipation. In the molecular kinetic theory, the advancing or receding contact lines have to overcome molecular energy barriers of the order of $k_BT$. In adaptation it is assumed that the solid surface spontaneously changes once it is in contact with the liquid; energy is released during this relaxation process as heat. In slide electrification charges are separated and electrostatic work has to be carried out. The respective contributions of the different processes and the related theories are, however, still under debate [16, 17]. One example is the role of surface defects. It is still debated if and how topological or chemical defects influence contact line motion and its velocity dependence [18].

Consequently, to date no quantitative theory can predict such a simple event as a drop sliding down a tilted plate. This lack of understanding is largely caused by a lack of combined spatial and temporal resolution in microscopical methods [19, 20]. To understand dynamic wetting and to distinguish between different molecular processes one would need to image the motion of contact lines with a resolution much better than 100 nm at a high frame rate. Usually, in experiments the so-called macroscopic or 'apparent' contact angle is measured with a camera. From the videos the contour of the liquid surface is extrapolated to the point of intersection with the solid surface [21]. The method is fast but the resolution is of the order of 10 μm. Using confocal microscopy, a resolution of the order of below 1 μm can be obtained [22], but the method is slow and limited to low speeds. Scanning electron microscopy (SEM) can achieve a higher resolution but beam damage, low frame rate and the presence of vacuum limits its applicability. In environmental SEMs, water drops can be imaged but the resolution is not much better than in optical microscopy [23, 24]. Much better resolution can be achieved in atomic force microscopy but imaging is challenging, the frame rate is low and one can never be sure that the tip does not influence the shape of the liquid near the contact line [25].

X-ray phase contrast imaging (XPCI) or X-ray holography using synchrotrons as an X-ray source can in principle overcome these limits. Few studies have already used X-ray imaging to investigate wetting in for example microchannels [26, 27].

In this work we propose X-ray phase contrast imaging to compute high-resolution holograms of the dynamic receding contact angle with a spatial resolution of about 50 nm. Therefore, a moving glass fiber is pulled out of a liquid bath filled with aqueous glycerol mixtures. The use of X-rays instead of light microscopy promises an improved spatial resolution compared to common optical methods.

## Experimental

The measurements were taken at the 'Göttingen Instrument for Nanoimaging with X-rays' (GINIX), which is an endstation situated at the coherence applications beamline P10 at the PETRA III storage ring at Deutsches Elektronen-Synchrotron (DESY) in Hamburg [28, 29]. The GINIX endstation utilizes coherent X-ray radiation for propagation-based X-ray phase-contrast imaging for high-resolution imaging of the sample [30, 31, 32]. The resulting images are near-field diffraction patterns, or holograms. A schematic depiction of the technical components can be seen in Figure 1a.

The X-ray source was a 5 m long undulator. The X-rays passed a monochromator, selecting an X-ray energy of 7.9 keV. The monochromator was followed by slits and a set of mirrors in Kirkpatrick-Baez geometry, where the X-rays were focused in vertical and horizontal direction. In the focal point the beam spot size was about 300 nm x 300 nm. The waveguide, which was situated in the focal point, acted as a quasi-point source and a low-pass filter to reduce high-frequency artifacts in the holograms. The sample stage (Figure 1b) was situated slightly behind the focus in the diverging beam. Thus, the placement of the sample resulted in a field of view of about 160 μm in both horizontal and vertical direction in the final holograms. After illumination with the cone shaped beam, the X-rays were collected at a distance of 5 m. Images were detected using a Zyla sCMOS camera with a fiber-optic plate detector (2560 x 2160 pixels with a respective pixel size of 6.5 μm x 6.5 μm).

The sample consisted of a 17 μm thick glass fiber which was pulled vertically out of a liquid bath by a programmable motor with a defined velocity in a range between 0 mm/s and 5 mm/s. Hereby, a 2-phase bipolar stepper motor from Oriental Motor was used. The glass fiber was additionally secured in its position with two syringes, minimizing the horizontal movement of the fiber in relation to the beam. The liquid bath (customized Teflon box) was closed with a lid, which was hydrophobically coated with Glaco spray (Glaco Mirror Coat Zero). The lid had a hole of 1 cm in diameter. A liquid dome was formed in the hole by slightly overfilling the liquid bath. This reduction of the liquid surface area ensured that the slightly inclined X-ray beam would not be clipped at the edge of the bath while measuring the liquid surface and the meniscus as close as possible. As liquid we used Milli-Q water-glycerol mixtures of 30 wt%, 50 wt% and 80 wt% glycerol (98 % level of purity from Fisher scientific). The glass fibers (Hybon 2002 fibers by Nippon Electric Glass) were pre-characterized by scanning electron microscopy (SEM, FEI Nova 600 NanoLab) (Figure 1c). Before inserting the fibers into the setup, they were cleaned in toluene.

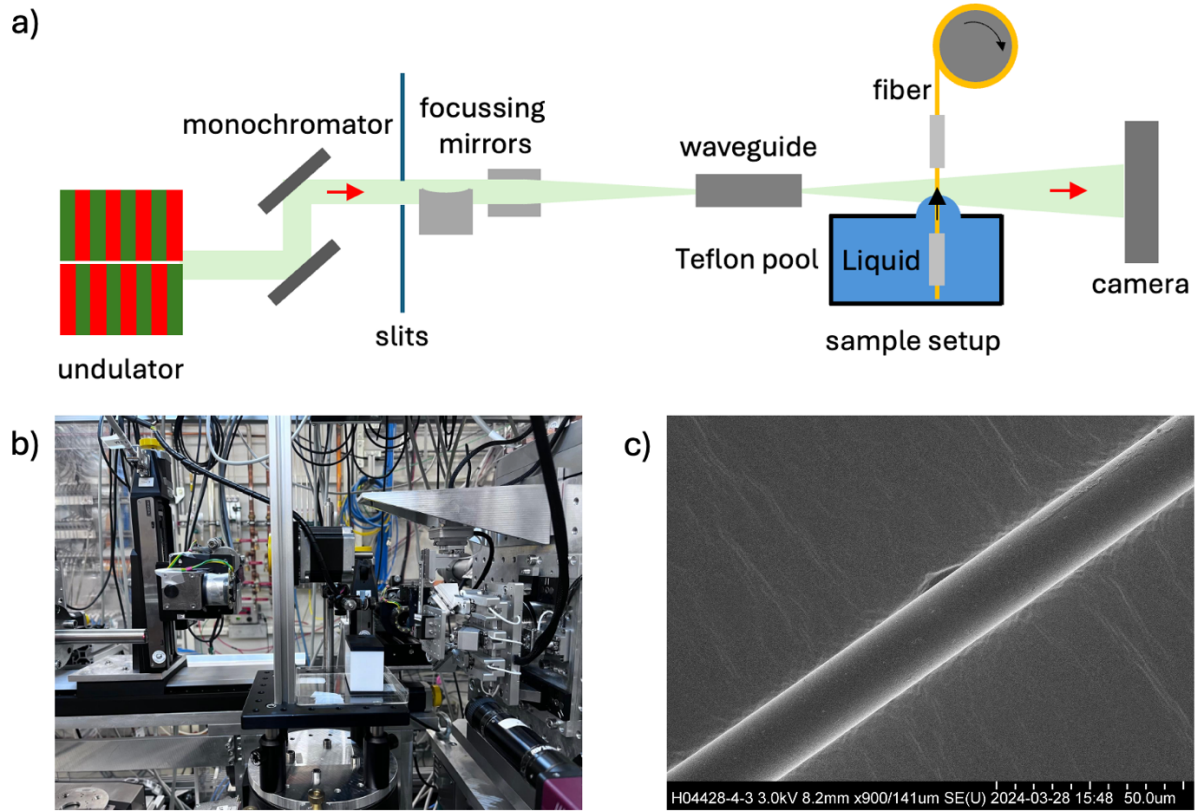

*Figure 1. a) A schematic depiction of the technical components of the GINIX endstation and the sample stage at the P10 beamline at DESY, Hamburg. b) A picture of the sample stage. c) An SEM image of a cleaned glass fiber of 17 μm thickness.*

## Results and Discussion

For each aqueous glycerol mixture and fiber speed we obtained holograms. A hologram of a fiber can be seen in Figure 2a. It shows the fiber and the meniscus of the liquid. In the respective example, a 80 wt % glycerol-water mixture has been used, and the fiber was pulled upward with a velocity of 1 mm/s. The air bubbles seen in the hologram were generated when mixing glycerol and water and further pouring the mixture into the liquid bath. However, the bubbles did not interfere with the liquid surface close to the fiber. As a result of the imaging method there is no further information on where exactly in relation to the fiber the bubbles were positioned as there is no depth of field in the holograms. We can exclude, however, that they were in the meniscus.

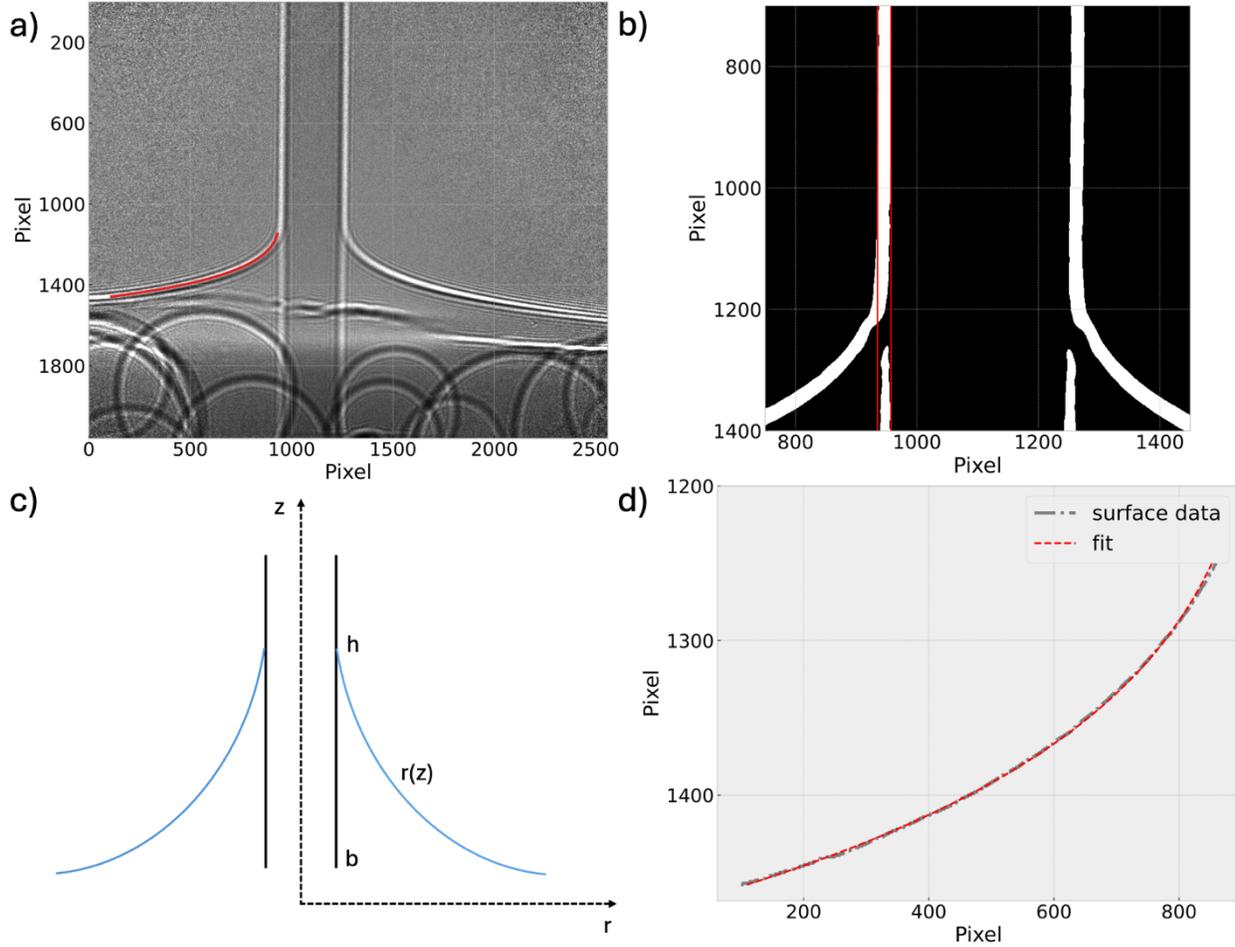

*Figure 2. a)* Hologram of a glass fiber pulled upward with a velocity of 1 mm/s in a 80 wt% glycerol-water mixture. The red line represents the fit of the liquid surface (see also d) *b)* A zoomed-in image of the glass fiber. The spatial error in the hologram is based on the extended fiber surface region, indicated by the two red lines. *c)* Schematic depiction of the meniscus on a vertical fiber including all parameters of the model described by eq. (1). *d)* Fit of the liquid surface.

The analysis was performed using the HoToPy Python-toolbox for XPCI [33]. To determine the spatial error in the obtained holograms the central fiber is magnified (Figure 2b). The two red lines indicate the fiber surface region, i.e., the point of change in contrast. In the obtained holograms we achieved a spatial resolution of 50 nm/pixel with a spatial error of 450 nm. This error is due to this extended surface region of the fiber, which amounts to more than one pixel in width. The exact placement of the real fiber surface within this region is ambiguous hence resulting in the spatial error of 450 nm which is half the length of the extended fiber surface region. The same spatial error could be observed for the interfacial region of the liquid surface.

The imaged liquid surface, or meniscus, on the vertical fiber can be described by solving the Laplace equation in radial coordinates [34]. In the holograms the field of view corresponds to a length of 160 μm. Within this regime, which is smaller than the capillary length, the capillary force surpasses the gravitational force, hence we neglected gravity in our model.

The meniscus can then be described using the following equation [35]:

$$r(z) = b \cosh\left(\frac{z-h}{b}\right), \quad (1)$$

where $h$ is the height measured at the three-phase contact point and $b$ is the radius of the fiber. A schematic depiction of the meniscus with all relevant parameters is displayed in Figure 2c.

The measured liquid surface can be fitted using the model described in equation (1) (Figure 2d). The shown data corresponds to the measurement of 80 wt% glycerol in Milli-Q water in Figure 2a. The model of the liquid meniscus sufficiently describes the curvature of the liquid surface. The quality of the fit can also be seen when the fit is plotted in the hologram (Figure 2a, red line).

Each measurement cycle for each solution and constant velocity consisted of the same sequence. First the X-ray beam was focused on a position where neither the glass fiber nor the liquid surface was visible, to collect empty background images. These are needed to compute the holograms, for which the intensity in sample image is divided by the intensity in the background image. Then, the beam position was changed to focus on the liquid meniscus and the glass fiber. In this configuration 200 images were taken. 50 images with a static, non-moving fiber (Figure 3a) followed by 100 images with a moving fiber (Figure 3b) and finally again 50 images with a static fiber after the motor was stopped. The exposure time for every image was 40 ms. Figures 3a and 3b show single holograms. The example shows data from a 80 wt% glycerol-water mixture, where in the dynamical case (Figure 3b) the glass fiber was pulled upwards with a velocity of 1 mm/s.

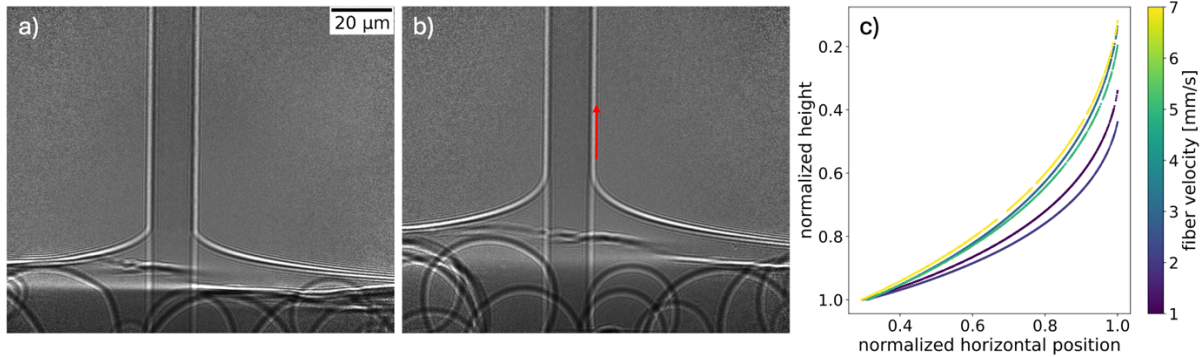

*Figure 3. Hologram series of a measuring cycle for 80 wt% glycerol in Milli-Q water: **a)** static fiber, **b)** moving fiber with a velocity of 1 mm/s. The red arrow indicates the direction of movement of the glass fiber. **c)** The liquid meniscus plotted for different fiber or contact line velocities for 80 wt% glycerol in Milli-Q water.*

The edges of the fiber as well as the liquid were extracted by scanning the image respectively horizontally and vertically and identifying the pixel positions corresponding to a jump in contrast (thresholding). The pixel positions associated to the liquid meniscus were fitted using the model described in equation (1). In Figure 3c, the curvature of the liquid meniscus is plotted for different fiber, i.e., contact line velocities ranging from 1mm/s (blue) to 7 mm/s (yellow). In order to directly compare the different curvatures, the baseline of the liquid surface was normalized and set to the same vertical position. The measured heights are hereby all divided by the value of the lowest vertical position of each curve, which corresponds to a higher pixel number. For this reason, when the real height increases, the value of the normalized height in the plot decreases. All curvatures start at the baseline value of 1. This is necessary to ensure the curvatures are visually comparable.

Within the 160 μm field of view the whole imaged liquid surface surrounding the glass fiber is pulled upward by the movement. The surface does not recede back to its equilibrium position within the field of view. The baseline of the liquid surface surrounding the fiber is therefore higher for higher contact line velocities.

Essentially the same correction is applied to account for horizontal movement of the fiber in between or during measuring cycles. The edge of the fiber surface is hereby set to the same position. The surface of the glass fiber was situated at the value of 1.

The highest point of the liquid meniscus, which is directly on the fiber surface, corresponds to the three-phase contact point. The highest vertical position of the three-phase contact point is reached for the highest velocity. The liquid is dragged upward with the glass fiber affecting the curvature of the meniscus (Figure 3c).

Three solutions with different glycerol fraction, and hence different viscosity have been measured (Figure 4). The analysis, i.e., the computation of the dynamic receding contact angle was done using single holograms of the moving or static fiber. These holograms were analyzed to determine the dynamic receding contact angle on the moving glass fiber. The angle was calculated within a region of 500 nm from the three-phase contact point by applying a tangent to the liquid surface in this region. For this tangential fit always 10 pixels were used. The dynamic contact angles can then be calculated using the angle or slope of the tangential fit of the liquid surface and the slope of the linear relation describing the fiber.

The dynamic receding contact angles tended to decrease with increasing speed of the contact line or fiber (Figure 4). This could be observed for all three aqueous glycerol mixtures. For each respective fiber speed five different holograms were analyzed. The dynamic receding contact angle on the glass fiber was then taken to be the mean value averaging over those five measurements. The error is the standard deviation of this value. The contact angle calculations were done for measurements in a velocity range between 0 mm/s, corresponding to a static fiber, and 5 mm/s.

The biggest change in the value of the dynamic receding contact angle was measured between the static glass fiber and the slowly moving fiber with a contact line velocity of 1 mm/s. This is again true for all three mixtures.

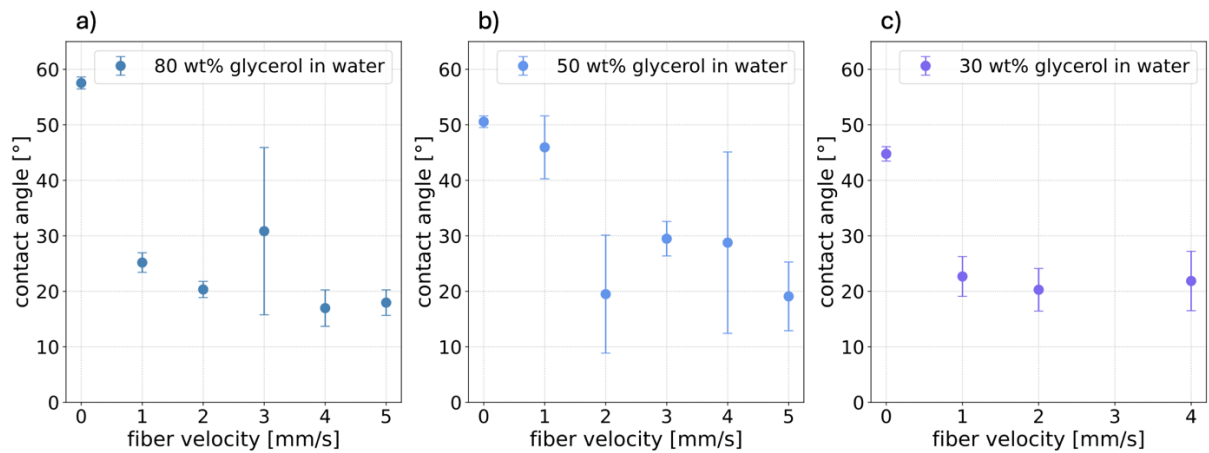

*Figure 4.* The dynamic receding contact angle in dependence of the fiber velocity for *a)* 80 wt%, *b)* 50 wt% and *c)* 30 wt% glycerol in Milli-Q water.

The largest contact angle of around 58° for a static glass fiber was measured for the mixture of 80 wt% glycerol in water. The angle decreased with a decreasing wt% of glycerol in water, hence decreasing viscosity. Large standard deviations, i.e., errors in the contact angle analysis stem from horizontal movement and vibrations of the fiber. The vibrations were mostly caused by the setup itself. The horizontal movement of the fiber was within the small range of about half the field of view, about 80 µm, because the syringes allowed some movement space. Both influence the overall quality of the holograms. Horizontal fiber movement as well as vibrations could be especially observed for larger velocities, also those exceeding 5 mm/s. Those velocities are therefore not presented in the analysis above.

## Conclusion

X-ray phase contrast imaging was demonstrated to be a feasible method to investigate dynamic wetting phenomena. Using this method, we could obtain holograms of a static and dynamic contact line by pulling a 17 µm thick glass fiber out of a liquid bath. Hereby, three different glycerol-water mixtures with varying viscosity have been studied. When the fiber is pulled upward, the entire liquid meniscus is elevated within the field of view of 160 µm. The shape of the meniscus is usually determined by the equilibrium between capillary forces and gravitational forces. Since our field of view was much smaller than the capillary length, gravity can be neglected. The liquid surface in the holograms can be fitted with the corresponding Laplace equation. The dynamic receding contact angle decreases with increasing contact line velocity for all three mixtures. We were able to determine the dynamic receding contact angle on moving glass fibers within a region of 500 nm measured from the three-phase contact point. In the holograms we achieved a spatial resolution of 50 nm/pixel with a spatial error of 450 nm at 1 – 5 mm/s contact line velocity. This experiment was the first dynamic wetting experiment, in which a moving contact line was imaged using X-ray phase contrast imaging. The method can be used in future experiments involving dynamic wetting phenomena, higher temporal and spatial resolution could potentially be achieved using an X-ray free electron laser.

## Associated Content

Supporting information contains information (graphical form) on data analysis workflow (PDF).

## Author Information

**Author Contributions**

The manuscript was written through contributions of all authors. All authors have given approval to the final version of the manuscript.

**Corresponding Author**

\* amannk@mpip-mainz.mpg.de

## Acknowledgements

The authors acknowledge DESY (Hamburg, Germany), a member of the Helmholtz Association HGF, for the provision of experimental facilities. Parts of this research were carried out at The GINIX endstation at beamline P10 of PETRA III. Beamtime was allocated for proposal I-20231182. This research was supported in part through the Maxwell computational resources operated at DESY. J.L. and T.S. are members of the Max Planck School of Photonics. T.E. acknowledges funding by the Centre for Molecular Water Science (CMWS) within Early Science Projects. We also acknowledge the scientific exchange and support of the CMWS. This work is partially supported by the German Research Foundation (DFG) within the framework Collaborative Research Centre 1194 ''Interaction of Transport and Wetting Processes'', Project-ID 265191195, subproject T02 (F. B.). We thank Stefan Geiter as well as the electronic workshop from the Max Planck Institute for Polymer Research for their technical support.

# Supporting Information

# Data analysis workflow

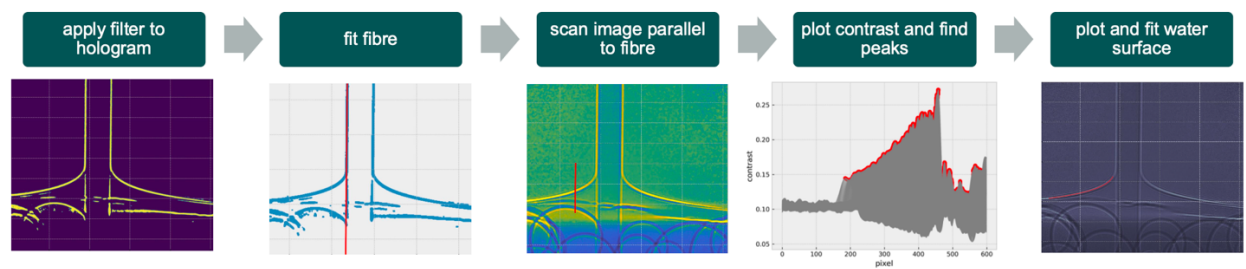

***Figure S.1.*** *Graphical representation of the image processing of a single hologram shown on an example for 80 wt% glycerol in water. First a filter is applied to the hologram to identify the interfacial information. In a next step the fiber is fitted using a linear relation. The surface area is scanned parallel to the fiber and the contrast peaks are identified. The positions of the highest contrast adhere to the positions of the liquid surface in the image. Finally, the liquid surface is fitted.*